\documentclass[10pt,oneside,twocolum,final]{IEEEtran}

\usepackage{amsmath,amsfonts}
\usepackage{amssymb}
\usepackage{algorithm}
\usepackage{algpseudocode}
\usepackage{bm}
\usepackage{booktabs}
\usepackage{array}
\usepackage[caption=false,font=scriptsize,labelfont=sf,textfont=sf]{subfig}
\usepackage{extpfeil}
\usepackage{textcomp}
\usepackage{stfloats}
\usepackage{url}
\usepackage{verbatim}
\usepackage{graphicx}
\usepackage{cite}
\usepackage{color}
\allowdisplaybreaks[4]

\begin{document}


\title{Spherical Antenna Arrays for Future Communications: Principles, Applications, and Research Directions}
\IEEEoverridecommandlockouts

\author{Cunhua~Pan,~\emph{Senior~Member},~\emph{IEEE},
	    Xianzhe~Chen,
        Hong~Ren,~\emph{Member},~\emph{IEEE},
        and~Jiangzhou~Wang,~\emph{Fellow},~IEEE

\thanks{\emph{(Corresponding author: Hong Ren and Jiangzhou Wang)}}

\thanks{C. Pan, X. Chen, H. Ren and Jiangzhou Wang are with the National Mobile Communications Research Laboratory, School of Information Science and Engineering, Southeast University, Nanjing 211189, China (e-mail: chen.xianzhe, hren, cpan, j.z.wang@seu.edu.cn). (Cunhua Pan and Xianzhe Chen are co-first authors.)}




}

\maketitle

\newtheorem{lemma}{Lemma}
\newtheorem{theorem}{Theorem}
\newtheorem{remark}{Remark}
\newtheorem{corollary}{Corollary}
\newtheorem{proposition}{Proposition}

\begin{abstract}
	With the development of 6G technologies, traditional uniform linear arrays (ULAs) and uniform planar arrays (UPAs) can hardly meet the demands of three-dimensional (3D) full-space coverage and high angular resolution. Spherical antenna arrays (SAAs), with elements uniformly distributed on a spherical surface, provide an effective solution. This article analyzes the issues of traditional arrays, summarizes the advantages and typical structures of SAAs, discusses their potential application scenarios, and verifies their superiority over UPAs via a case study. Finally, key technical challenges and corresponding research directions of SAAs are identified, providing a reference for their research and application in future wireless communications.

\end{abstract}

\section{Why Need Spherical Antenna Arrays?}\label{SA_Section1}

With the rapid evolution of wireless communication technologies toward 6G, particularly in key application scenarios such as integrated sensing and communication (ISAC) and the low-altitude economy, antenna arrays are required to meet increasingly challenging demands in terms of three-dimensional (3D) coverage, angular resolution, hardware complexity, and environmental adaptability \cite{10054381}. As a core component of wireless communication systems, the geometric configuration of an antenna array directly determines its signal radiation and reception performance, as well as overall system efficiency. Traditional uniform linear arrays (ULAs) and uniform planar arrays (UPAs) are inadequate to address the multi-dimensional communication requirements of complex scenarios. Against this background, spherical antenna arrays (SAAs), due to their unique spatial geometry, are anticipated to become a research focus in wireless communications, offering a novel technical pathway to overcome the performance bottlenecks inherent in conventional array structures.

\subsection{Bottlenecks of Linear and Planar Arrays}\label{SA_Section2}  

Traditional ULAs and UPAs exhibit inherent limitations that severely constrain system performance in practical wireless communication applications. ULAs support beam scanning in only one dimension \cite{liu_letter1}, resulting in limited angular resolution and narrow spatial coverage, which renders them incapable of addressing multi-directional or dynamically varying communication targets in 3D space. Although UPAs enable two-dimensional beam steering, they suffer from substantial performance degration, with angular resolution decreasing significantly as the target departs from the boresight direction \cite{9226455,11427011,jiang20263d}. In 3D scenarios such as low-altitude unmanned aerial vehicle (UAV) swarms, this often leads to issues including signal outage, mutual coupling, and coverage holes. Furthermore, improving the resolution of conventional arrays typically requires to increase the array aperture and the number of elements, which in turn substantially increases the complexity of radio frequency (RF) chains and phase shifters, thereby increasing system power consumption, cost, and physical size \cite{9745122}. Such trade-offs conflict with the application requirements of miniaturized platforms such as low-altitude terminals and satellites. Collectively, these limitations render traditional arrays unsuitable to meet the demands of future wireless communication applications, including ISAC, UAV swarm networking, and satellite communications \cite{1613527,10005237}.

\begin{figure}[t]
	\centering
	\includegraphics[scale=0.5]{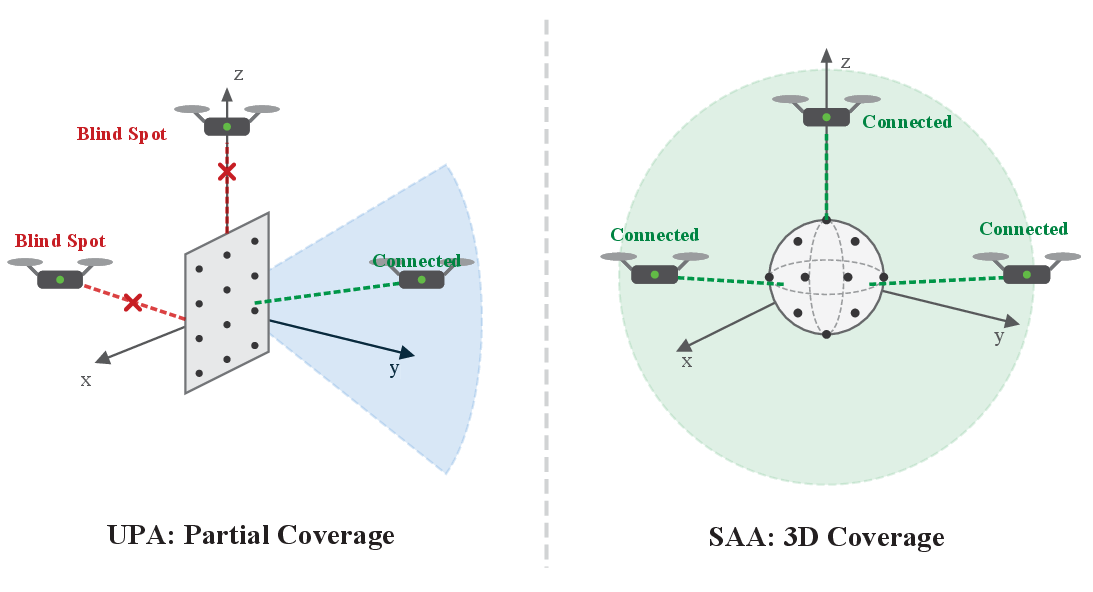}\\  
	\caption{Beam comparison}\label{p_comparison} 
\end{figure}

\begin{figure*}[t]
	\centering
	\includegraphics[scale=0.45]{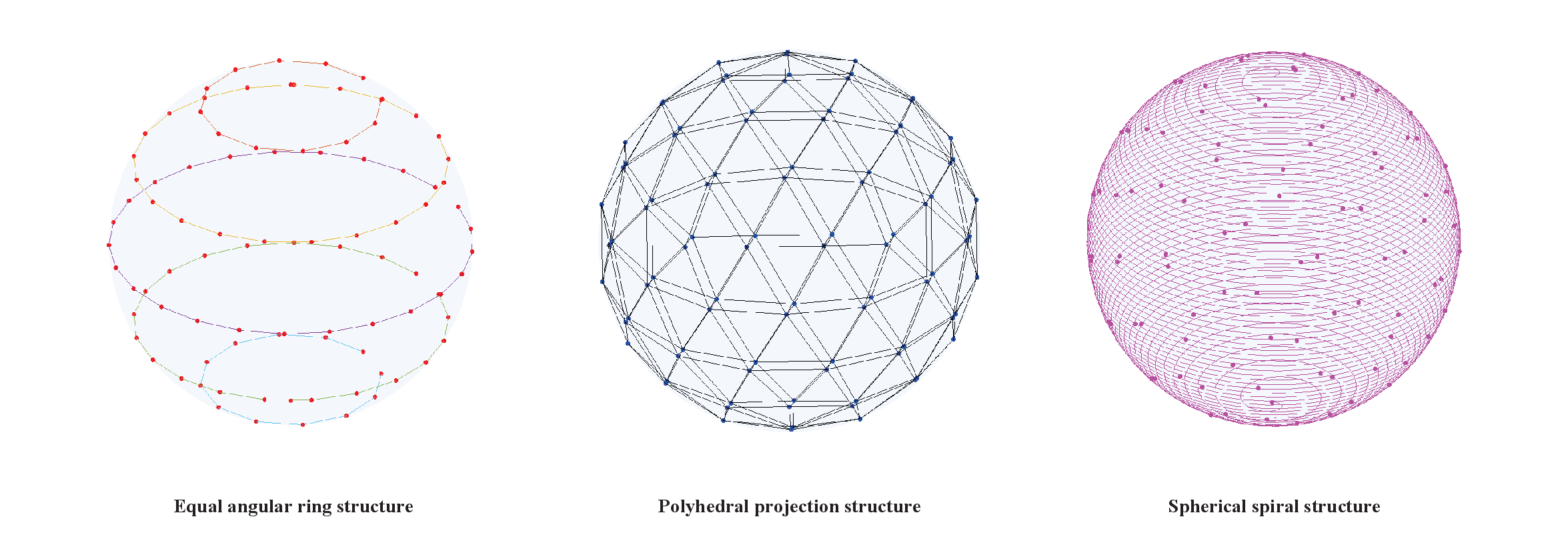}\\ 
	\caption{Typical Structures of Spherical Antenna Arrays.}\label{fig_sim1}
\end{figure*}

\subsection{Advantages of Spherical Antenna Arrays}\label{Hybrid_Sec3}

An SAA is a type of 3D antenna array in which elements are uniformly distributed on a spherical surface, forming a symmetric spatial structure. Unlike conventional ULAs or UPAs confined to one or two spatial dimensions, SAAs leverage the inherent geometric properties of the sphere to enable multi-directional signal interaction, as shown in Fig.~\ref{p_comparison}. Owing to their 3D spherical geometry, SAAs offer a range of distinct performance advantages over traditional linear and planar arrays, making them a promising solution for overcoming the limitations of conventional array architectures.

First, the uniform distribution of elements on the spherical surface enables 360° full-space coverage without blind spots, allowing signal reception and transmission for targets in any direction without physical rotation \cite{9373344}. This effectively overcomes the coverage limitations and performance degration inherent to traditional arrays.

Second, SAAs provide more isotropic and superior 3D angular resolution. With appropriate element arrangements and beamforming techniques, they can achieve accurate multi-target identification and tracking under low hardware complexity, making them particularly suitable for scenarios with dense target distributions and wide angular variations \cite{6905568}.

Furthermore, the spatial geometric symmetry of SAAs provides them with enhanced anti-interference capability and signal stability. Flexible beam steering can be employed to suppress sidelobe signals and form nulls. In addition, the flexibility in element layout helps mitigate signal occlusion and mutual coupling, thereby further improving communication system performance \cite{10817230,10964237}.

\subsection{Typical Structures of Spherical Antenna Arrays} 

Several representative geometric structures of SAAs are illustrated in Fig.~\ref{fig_sim1}.
The equal angular ring structure arranges elements along multiple parallel circular rings at equal angular intervals. As the most classical configuration, it has the benefits of geometric regularity and analytical tractability \cite{10964237}.
The polyhedral projection structure maps the vertices of a subdivided regular polyhedron onto the spherical surface, yielding highly uniform element spacing and excellent spherical symmetry. This structure is particularly suitable for high-performance massive MIMO and spatial signal processing systems \cite{9373344}.
The spherical spiral structure places elements along a continuous spiral curve on the spherical surface, avoiding periodic structural repetition and enabling quasi-continuous spatial coverage. This configuration is advantageous for broadband beamforming and high-resolution direction estimation \cite{famoriji2021source}.

\begin{figure*}[t]
	\centering
	\includegraphics[scale=0.63]{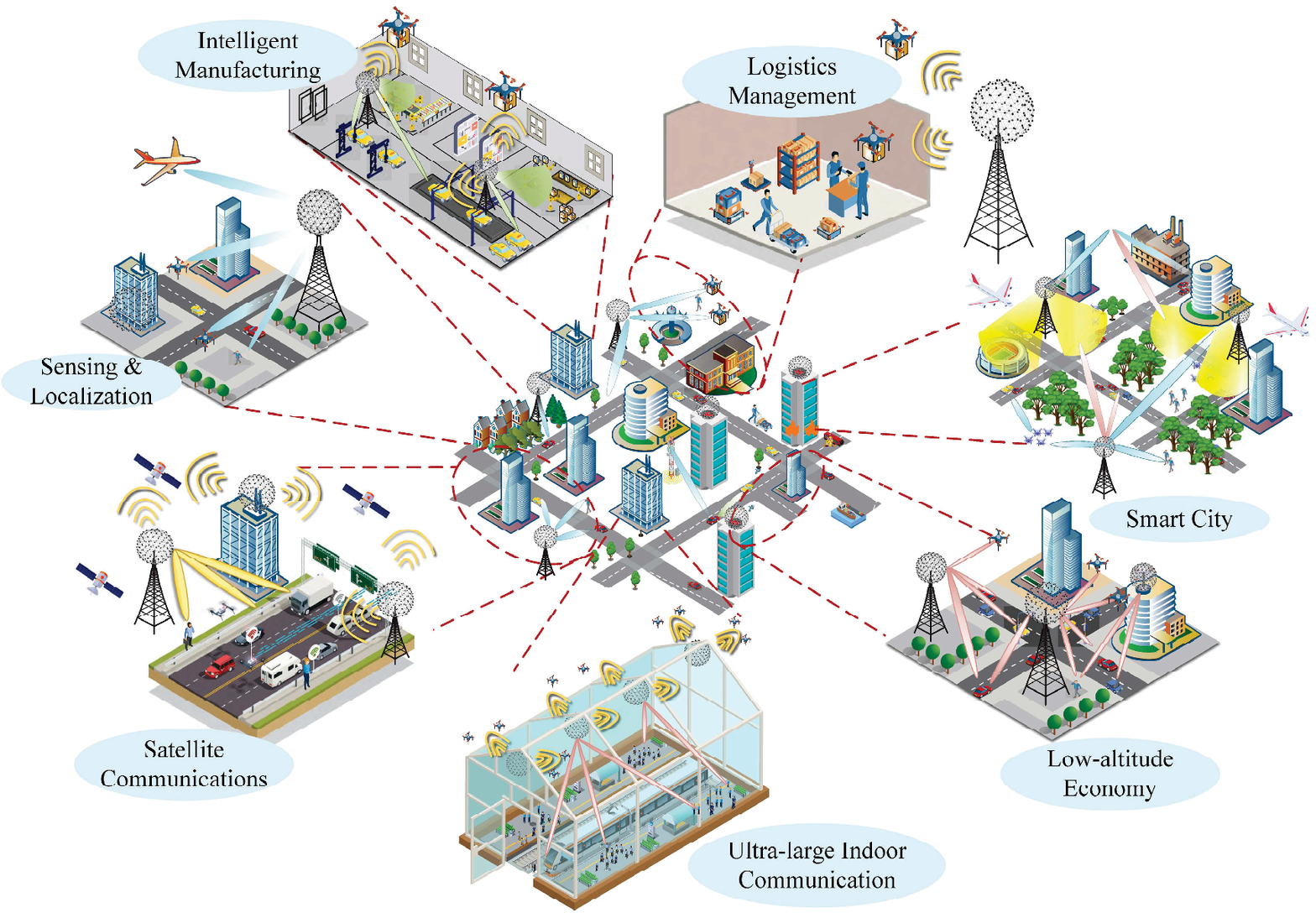}\\ 
	\caption{Potential application scenarios}\label{p_app} 
\end{figure*}

\section{Application Scenarios of Spherical Antenna Arrays}

With its unique spatial geometry, the SAA shows broad application prospects in wireless communications and related fields, as illustrated in Fig.~\ref{p_app}. The prospective scenarios are detailed below, with emphasis on the specific benefits offered in each context.

\subsection{Sensing and Localization}

Sensing and localization applications demand high precision in multi-target and complex propagation environments. Traditional ULAs and UPAs suffer from restricted spatial coverage and anisotropic performance, struggling to maintain consistent accuracy for scenarios such as indoor personnel localization and outdoor object detection, where targets may appear from any direction.

The spherical symmetry of SAAs can provide continuous covergae between boresight and edge directions, ensuring uniform angular resolution regardless of target angle. This isotropic sensing capability is critical for applications like indoor localization where user trajectory may be three-dimensional. Furthermore, the 3D structure enables simultaneous estimation of both azimuth and elevation angles using a single array, eliminating the need for multiple array nodes or mechanical scanning.

\subsection{Satellite Communications}

Satellite communication systems require wide coverage, high link reliability, and compact structures. Satellites across different orbits demand flexible beam alignment, while spaceborne and ground terminals face strict size, weight, and power constraints that render mechanically steered or conventional planar arrays unsuitable.

Unlike planar arrays that suffer from gain degradation at large scan angles, the spherical geometry maintains consistent gain in all directions. A single SAA-equipped BS can track satellites across Earth orbits without physical rotation, avoiding the link outage. For inter-satellite links, the ability to simultaneously form multiple beams enables a single array to maintain connections with multiple satellites across the entire satellite network.

\subsection{Ultra-large Indoor Communication}

Ultra-large indoor environments such as high-speed railway stations, airports, and large shopping malls feature expansive spaces, complex structures, high population density, and severe multipath interference. Conventional communication arrays often struggle to provide uniform coverage and reliable signal quality, resulting in blind spots, signal attenuation, and load imbalance.

Deployed on concourse ceilings, SAAs provide consistent coverage across large open spaces without the gain degradation at coverage edges that characterizes planar arrays. This is particularly valuable in railway stations where passengers are distributed across ticketing areas, waiting halls, and platforms, all requiring uniform service quality. The ability to form multiple directional beams allows a single SAA to serve high-density areas with narrow focused beams while maintaining wide coverage for sparse areas, effectively balancing load without multiple sectorized arrays.

\subsection{Low-altitude Economy}

The low-altitude economy, including UAV swarms, logistics, and tourism, requires advanced 3D communication and sensing capabilities. UAV swarms exhibit dynamic spatial distributions and rapidly varying spatial configurations, demanding real-time beam tracking across multiple moving targets, a challenge for traditional arrays constrained by limited coverage and mechanical scanning delays.

The 360° coverage of SAAs is particularly crucial in UAV swarm scenarios, where targets can appear in any directions of the array. SAAs can track drones flying trajectory while simultaneously maintaining links with ground control stations without antenna panel switching delays. The ability to form multiple independent beams allows a single ground-based SAA to serve an entire swarm, reducing infrastructure cost. For UAV-mounted terminals, the symmetric structure eliminates attitude alignment requirements, as communication links remain stable regardless of drone orientation. During takeoff and landing, where drones undergo rapid altitude changes and require uninterrupted connectivity for safe operation, the continuous full-sphere coverage ensures seamless signal connection without blind spots or link interruptions.

\subsection{Smart City}

Smart city construction spans intelligent transportation, security, and environmental monitoring, demanding ubiquitous coverage and high reliability in complex urban environments. Conventional antenna configurations often struggle to achieve seamless coverage and consistent performance across diverse deployment scenarios with high node density and complex signal propagation.

At urban intersections, the full-sphere coverage of SAAs enables simultaneous communication with vehicles on all approaching roads from a single deployment point, eliminating the need for multiple directional arrays. For security applications, a single SAA deployed on a building corner provides continuous surveillance across both streets without blind spots. In environmental monitoring, the uniform angular response reduces the complexity of calibration and data integration when aggregating sensor readings from scattered IoT devices across the urban landscape.

\subsection{Intelligent Manufacturing}

Intelligent manufacturing environments are characterized by dense deployments of sensors, robotic arms, and automated guided vehicles, requiring reliable, low-latency wireless connectivity. Industrial environment suffers complex propagation conditions with significant multipath effects, signal blockage from metallic equipment, and electromagnetic interference, posing considerable challenges for conventional arrays.

The omnidirectional nature of SAAs is advantageous in factory floors where  metal shelving and equipment cause complex signal shadowing. Unlike directional arrays that may lose connectivity when a robotic arm moves between the array and a target, SAAs provide multiple spatial paths for signal propagation, increasing link robustness. For automated guided vehicles traveling through confined corridors, connectivity remains stable regardless of vehicle orientation, eliminating the need for complex attitude tracking systems.

\subsection{Logistics Management}

Modern logistics management involves warehouse inventory tracking, fleet coordination, and last-mile delivery, requiring continuous, reliable connectivity across large-scale indoor and outdoor spaces. Traditional infrastructures face coverage limitations in warehouse environments with metal shelving and in dynamic outdoor scenarios where terminal locations vary frequently.

In warehouse settings where metal racking creates signal shadows, the 360° coverage of SAAs ensures that inventory tags and automated guided vehicles remain connected regardless of aisle or facing direction, eliminating the need for dense access point deployment. For outdoor fleet management, a single SAA at a distribution center can simultaneously track vehicles arriving from all directions and communicate with delivery drones, providing unified connectivity that would otherwise require multiple specialized systems.

\section{Case Study}\label{SA_Section6}

To validate the superior performance and practical advantages of SAAs in 3D space signal transmission, this section presents a comparative case study between the SAA architecture and the conventional UPA.

\begin{figure}[t]
	\centering
	\includegraphics[scale=0.68]{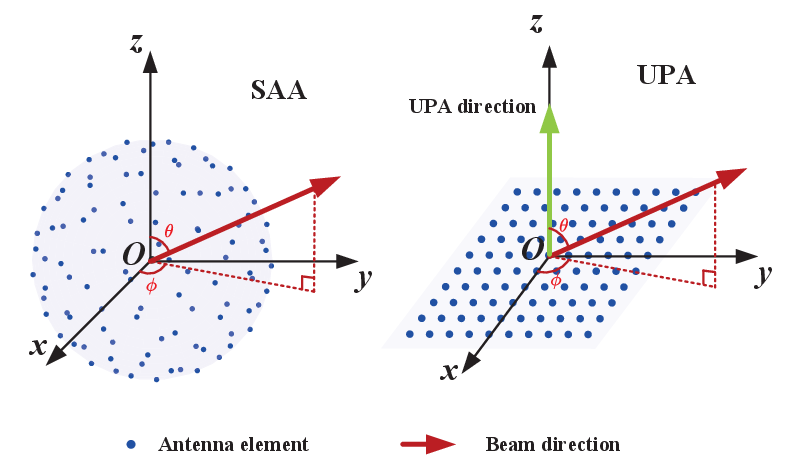}\\ 
	\caption{SAA and UPA}\label{p_spherical} 
\end{figure}

As illustrated in Fig.~\ref{p_spherical}, the SAA comprises $N$ elements deployed on the surface of a sphere with radius $R$, where a golden-angle spiral distribution is adopted to ensure uniform spatial coverage. The coordinates of each element are determined based on the golden angle, a specific angular value that enables the uniform spiral arrangement across the sphere’s surface.
By contrast, the UPA is arranged on the $xy$-plane as a $\sqrt{N} \times \sqrt{N}$ array with half-wavelength element spacing.

For the analysis purpose, only line-of-sight (LoS) channels are considered for both array architectures. For the SAA, LoS channels are only present for elements facing the target direction. In contrast, LoS channels for the UPA are restricted to the forward $+z$ hemisphere.

The channel coefficient for each array element is modeled based on the propagation distance from the element to the target, accounting for signal attenuation and phase shift during propagation.
Conjugate beamforming weights are obtained by normalizing the desired channel vector, which ensures effective beam focusing toward the target direction. The normalized beam power is used to evaluate the focusing performance of the beam formed by each array.

\subsection{Angular Beam Pattern}

\begin{figure}[t]
	\centering
	\includegraphics[scale=0.405]{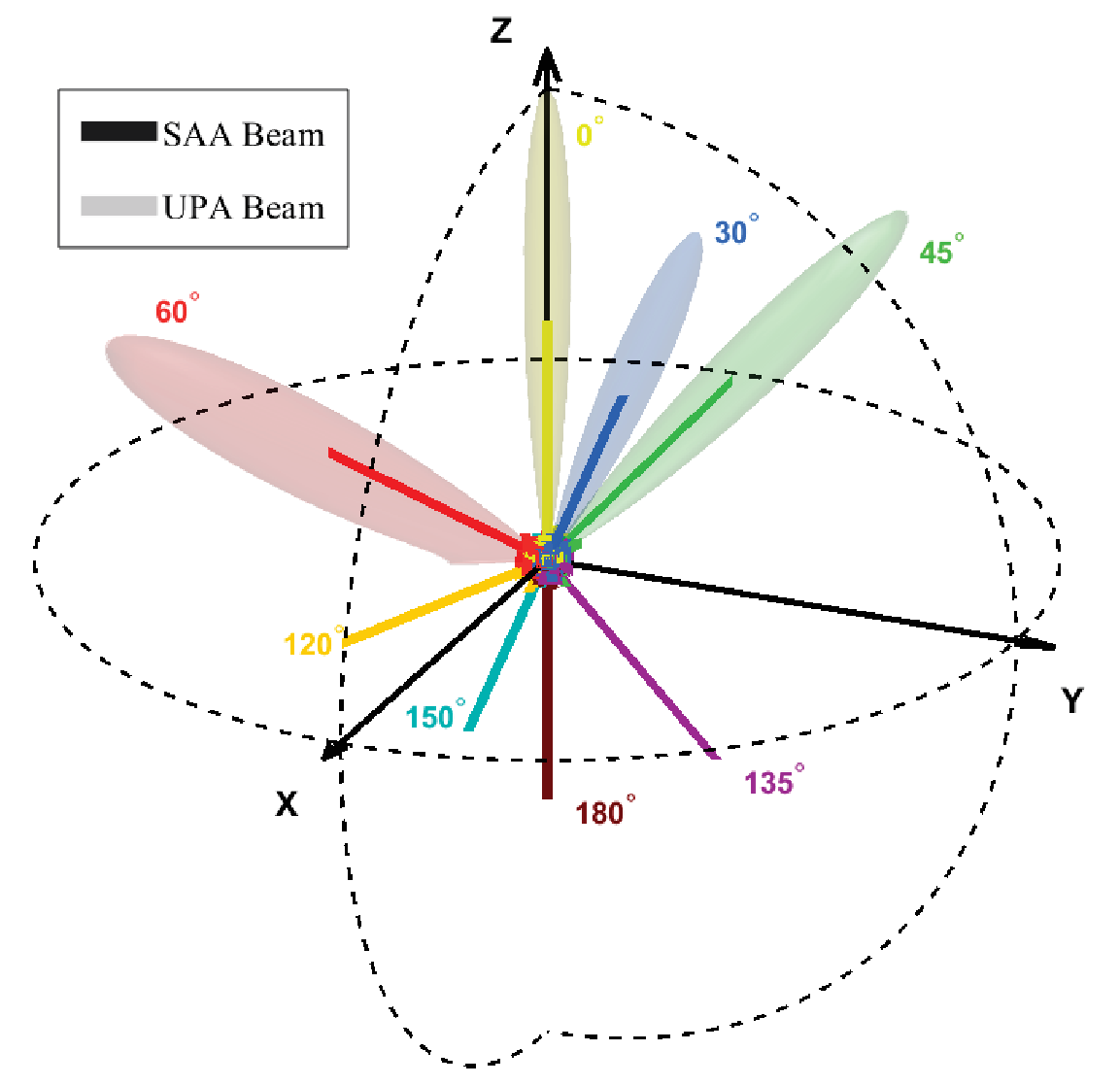}\\ 
	\caption{Angular Beam Patterns with Eight Focal Points}\label{p_beam_angle}
\end{figure}

The angular beam patterns of the SAA and UPA are illustrated in Fig.~\ref{p_beam_angle}, with key simulation parameters detailed as follows. 
The radius of the SAA is set to \( R = 0.5 \, \text{m} \), 
while the element spacing of the UPA is \( d = \lambda/2 \). 
Both arrays are configured with \( N = 100 \) antenna elements, 
operating at a wavelength of \( \lambda = 0.01 \, \text{m} \). 
The focal distance is fixed at \( r_0 = 30 \, \text{m} \). 
Eight focal points are adopted for performance evaluation, with azimuth and elevation angles specified as:  
$(\phi, \theta)$ $=$ $(\pi/6, \pi/6)$, $(5\pi/6, \pi/3)$, $(\pi/3, \pi/4)$, $(3\pi/4, 2\pi/3)$, $(\pi/4, 3\pi/4)$, $(2\pi/3, 5\pi/6)$, $(\pi/3, 0)$, $(\pi/3, \pi)$.
The beam patterns are assessed over the full angular range, i.e., \( \theta \in [0, \pi] \) (elevation) and \( \phi \in [0, 2\pi] \) (azimuth), with \( 181 \times 181 \) sampling points used for discrete evaluation.



As observed from Fig.~\ref{p_beam_angle}, the SAA maintains excellent and stable beam focusing performance in the full elevation range from \(0^\circ\) to \(180^\circ\). Regardless of how the elevation angle of the focal point changes, the beamwidth, peak gain and side-lobe level of the SAA remain almost unchanged, showing strong wide-angle adaptability and spatial isotropy.

On the contrary, the UPA suffers from serious beam performance degradation with the increase of elevation angle. When the focal point is close to \(90^\circ\), the beamwidth of the UPA is significantly broadened, the directivity is sharply reduced, and the focusing accuracy is seriously deteriorated. More critically, when the elevation angle of the target exceeds \(90^\circ\), the UPA cannot form an effective directional beam towards the target direction, which means that the UPA cannot provide reliable communication or sensing services for nearly half of the spherical space. This obvious performance gap fully proves that the SAA has overwhelming advantages in full-spherical 3D coverage and wide-angle beamforming compared with the traditional planar array.

\subsection{Distance Beam Pattern}

\begin{figure}[t]
	\centering
	\includegraphics[scale=0.41]{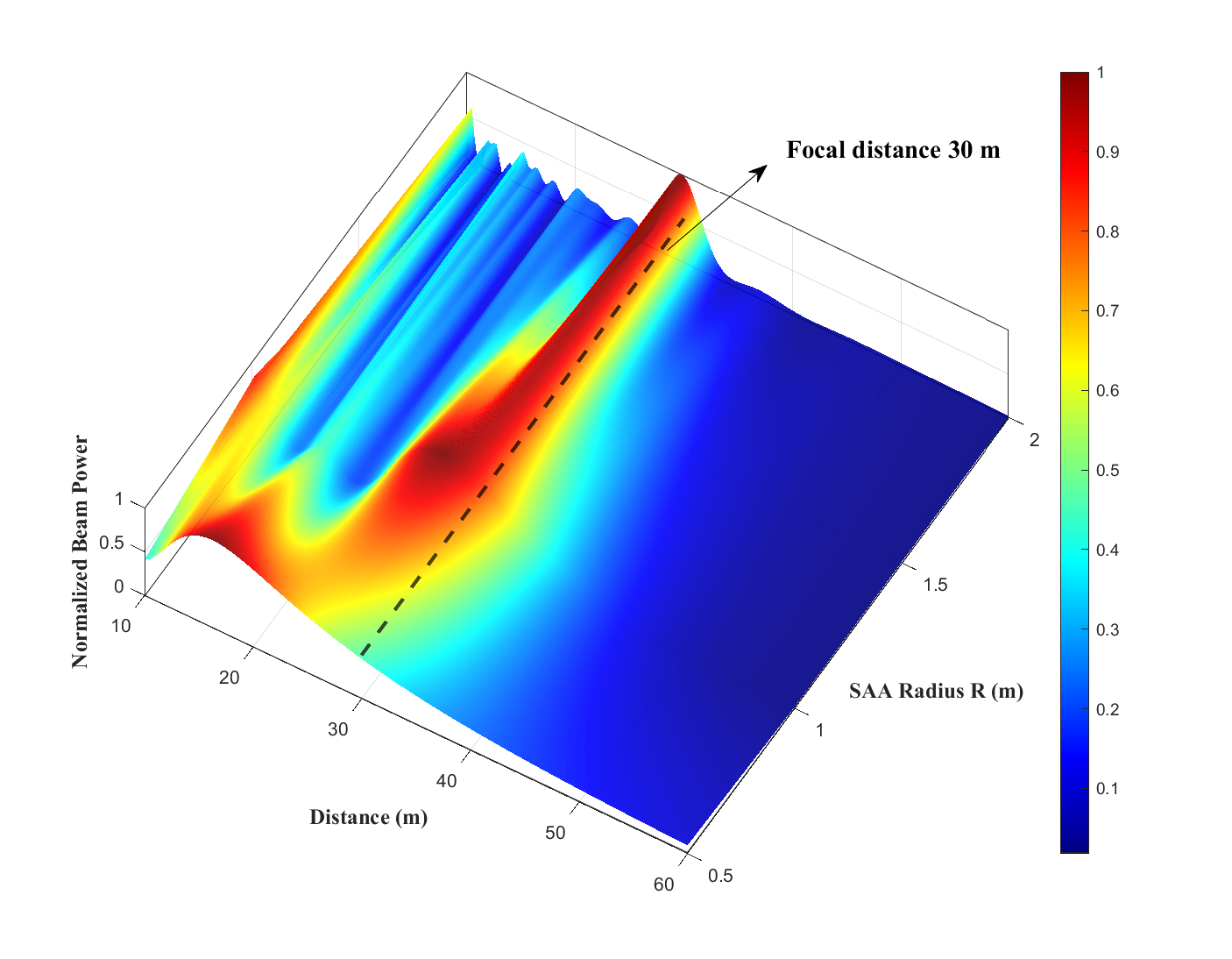}\\ 
	\caption{Distance Beam Patterns}\label{p_beam_distance} 
\end{figure}

Fig.~\ref{p_beam_distance} shows the normalized beam power distribution of SAAs with different radii along the distance dimension. 
It is observed that, similarly to the UPA, the SAA also possesses near-field distance-focusing capability. As the radius \(R\) of the SAA increases, the distance-focusing effect is significantly enhanced: the main beam becomes sharper, the side-lobe levels decrease, and the energy concentration improves. Notably, when the radius reaches \(R = 2 \, \text{m}\), the SAA achieves precise focusing at the target distance of 30 m, exhibiting strong range selectivity and high focusing resolution.
It should also be noted that the distance-focusing capability of SAAs can be further improved by employing higher operating frequencies, such as those in the terahertz (THz) bands. In fact, the distance-focusing capability of an SAA is directly related to the ratio of the array radius to the operating wavelength: a larger ratio yields a stronger distance-focusing capability.

Benefiting from the inherent spatial symmetry of the spherical structure, the distance-focusing performance of the SAA will not be reduced or deteriorated with the change of the azimuth or elevation angle, which is difficult to achieve by the planar array. These characteristics indicate that the SAA has unique and irreplaceable advantages in high-precision near-field positioning, target detection, 3D adaptive beamforming and spatial signal transmission applications, and is expected to become a key array technology in the next generation of wireless communication systems.

\section{The Relevant Challenges and Research Directions}\label{SA_Section7}

Despite the promising advantages of SAAs, their practical deployment faces critical technical challenges in channel measurement and modeling, channel estimation, codebook feedback, beam training and scanning, and sensing and localization. Addressing these issues requires systematic research integrating electromagnetics, signal processing, and machine learning. This section outlines key challenges and potential research directions.

\subsection{Channel Measurement and Modeling}

Accurate channel modeling for SAAs faces three main challenges. Existing measurement approaches designed for planar or linear arrays cannot be directly extended to 3D spherical geometries, making it difficult to characterize mutual coupling and curvature-dependent radiation. Current channel models also lack integration with SAA-specific characteristics such as spherical phase curvature and aperture projection effects. Moreover, emerging applications like UAV swarms and high-speed railways demand dynamic channel models that capture spatial and temporal dynamics, which time-invariant models cannot provide.

Future research should develop specialized 3D experimental platforms with integrated calibration procedures, supported by virtual antenna array technology to reduce hardware costs. Promising directions include constructing 3D parameterized sparse channel models incorporating spherical harmonic expansion and ray tracing, as well as developing machine learning-based dynamic modeling approaches for adaptive parameter updates.

\subsection{Channel Estimation}

Channel estimation for SAAs is hindered by high dimensionality, leading to excessive pilot overhead and reduced spectral efficiency. Non-ideal conditions such as mutual coupling, RF chain distortions, and hardware impairments further degrade accuracy. Near-field estimation is particularly challenging, as spherical wavefronts require joint angle-distance estimation.

Promising research directions include two-stage estimation frameworks that exploit channel parameter time invariance and common information among users, reducing pilot overhead to scale with sparse propagation paths. Modified array manifold matrices incorporating non-ideal factors should be developed, with spherical harmonic decomposition enabling mutual coupling compensation. For near-field operation, parameterized sparse channel models combined with sparse Bayesian learning offer high-precision joint angle-distance estimation.

\subsection{Codebook Feedback}

Efficient codebook feedback for SAAs faces several challenges. DFT-based codebooks developed for planar arrays are incompatible with the 3D curved manifold of SAAs, causing significant quantization errors. Full-space coverage combined with near-field distance dimensions leads to prohibitively large codebook sizes and excessive feedback overhead. Additionally, protocol-compatible feedback mechanisms under limited bit budgets require effective mapping of the SAA manifold to standard virtual array bases.

Future efforts should focus on SAA-specific 3D codebooks based on geodesic polyhedron structures and spherical harmonic expansion, aligning code vectors with physical element directions while incorporating near-field focusing capabilities. Array manifold separation techniques can facilitate mapping curved manifolds to DFT bases. Deep learning combined with sparse coding offers promising approaches for codebook compression, and hierarchical coarse-to-fine codebooks can balance coverage with feedback efficiency.

\subsection{Beam Training and Scanning}

Beam training and scanning for SAAs face multiple performance constraints. Full-space 3D beam scanning suffers from excessive latency due to exhaustive search, and existing 2D algorithms cannot meet 3D requirements. The near-field regime adds complexity as distance dimensions exponentially expand codebook scales. Hardware constraints on miniaturized platforms and array imperfections further degrade performance, while ISAC applications lack joint multi-beam training algorithms.

Promising solutions include low-overhead scanning schemes integrating lightweight neural networks. Hierarchical coarse-to-fine scanning strategies can substantially reduce search latency. For near-field operation, angle-distance joint focusing codebooks and training algorithms based on sparse sampling are essential. Hardware constraints can be addressed through hybrid precoding architectures with platform motion compensation. For ISAC applications, multi-beam joint training algorithms leveraging SAA symmetry can achieve highly separated multi-beam formation.

\subsection{Sensing and Localization}

Sensing and localization with SAAs face distinct challenges. Near-field 3D sensing models are lacking, as traditional far-field planar wave algorithms cannot be applied directly. Achieving high-precision joint angle-distance positioning involves complex trade-offs between algorithm complexity and computational constraints on miniaturized platforms. Robustness in complex environments with multi-path, non-line-of-sight propagation, and dense targets remains challenging, and theoretical benchmarks such as the Cramér-Rao bound for near-field scenarios are absent.

Future research should construct near-field spherical wave signal models incorporating distance parameters and derive array-target geometric relationships for 3D positioning. High-precision low-complexity algorithms can be developed through spherical harmonic decomposition, high-order pseudo-intensity vector techniques, and lightweight machine learning. Algorithms incorporating spatial filtering and sparse reconstruction for multi-path suppression should be developed, along with multi-target joint sensing for dense scenarios. Deriving the Cramér-Rao bound for SAA near-field sensing would provide theoretical accuracy limits to guide system design.

\section{Conclusion}\label{SA_Section8}


This article has demonstrated that SAAs offer unique advantages over conventional ULAs and UPAs, including 360° full-space coverage, isotropic angular resolution, and superior near-field focusing capability. A comparative case study confirmed that SAAs maintain stable beamforming performance across all directions, while planar arrays fail to cover the backward hemisphere. These features make SAAs attractive for emerging 6G applications such as integrated sensing and communication, satellite communications, and low-altitude economy. Despite these benefits, practical deployment faces critical challenges in channel modeling, estimation, codebook design, beam training, and localization. Addressing these issues through interdisciplinary research will position SAAs as a key enabling technology for future wireless communication systems.

\bibliographystyle{IEEEtran}
\bibliography{IEEEabrv,Reference}

\begin{thebibliography}{10}
\providecommand{\url}[1]{#1}
\csname url@samestyle\endcsname
\providecommand{\newblock}{\relax}
\providecommand{\bibinfo}[2]{#2}
\providecommand{\BIBentrySTDinterwordspacing}{\spaceskip=0pt\relax}
\providecommand{\BIBentryALTinterwordstretchfactor}{4}
\providecommand{\BIBentryALTinterwordspacing}{\spaceskip=\fontdimen2\font plus
\BIBentryALTinterwordstretchfactor\fontdimen3\font minus
  \fontdimen4\font\relax}
\providecommand{\BIBforeignlanguage}[2]{{%
\expandafter\ifx\csname l@#1\endcsname\relax
\typeout{** WARNING: IEEEtran.bst: No hyphenation pattern has been}%
\typeout{** loaded for the language `#1'. Using the pattern for}%
\typeout{** the default language instead.}%
\else
\language=\csname l@#1\endcsname
\fi
#2}}
\providecommand{\BIBdecl}{\relax}
\BIBdecl

\bibitem{10054381}
C.-X. Wang, X.~You, X.~Gao, X.~Zhu, Z.~Li, C.~Zhang, H.~Wang, Y.~Huang,
  Y.~Chen, H.~Haas, J.~S. Thompson, E.~G. Larsson, M.~D. Renzo, W.~Tong,
  P.~Zhu, X.~Shen, H.~V. Poor, and L.~Hanzo, ``On the road to {6G}: Visions,
  requirements, key technologies, and testbeds,'' \emph{IEEE Commun. Surveys
  Tuts.}, vol.~25, no.~2, pp. 905--974, 2nd Quart. 2023.

\bibitem{liu_letter1}
W.~Liu, H.~Ren, C.~Pan, and J.~Wang, ``Deep learning based beam training for
  extremely large-scale massive {MIMO} in near-field domain,'' \emph{IEEE
  Commun. Lett.}, vol.~27, no.~1, pp. 170--174, Jan. 2023.

\bibitem{9226455}
J.~Wang, W.~Deng, X.~Li, H.~Zhu, M.~Nair, T.~Chen, N.~Yi, and N.~J. Gomes,
  ``{3D} beamforming technologies and field trials in {5G} massive {MIMO}
  systems,'' \emph{IEEE Open J. Veh.Technol.}, vol.~1, pp. 362--371, 2020.

\bibitem{11427011}
X.~Chen, H.~Ren, C.~Pan, C.-X. Wang, and J.~Wang, ``Exploring the advantages of
  sparse arrays in near-field xl-mimo systems: Beam analysis and {EDoF}
  function,'' \emph{IEEE Trans. Commun.}, 2026, early access.

\bibitem{jiang20263d}
\BIBentryALTinterwordspacing
H.~Jiang, Z.~Dong, Z.~Zhou, and Y.~Zeng, ``{3D} spherical directly-connected
  antenna array for low-altitude {UAV} swarm {ISAC},'' 2026. [Online].
  Available: \url{https://arxiv.org/abs/2603.17620}
\BIBentrySTDinterwordspacing

\bibitem{9745122}
M.~Mahmood, A.~Koc, and T.~Le-Ngoc, ``3-{D} antenna array structures for
  millimeter wave multi-user massive {MIMO} hybrid precoder design: A
  performance comparison,'' \emph{IEEE Commun. Lett.}, vol.~26, no.~6, pp.
  1393--1397, Jun. 2022.

\bibitem{1613527}
B.~Tomasic, J.~Turtle, and S.~Liu, ``Spherical arrays - design
  considerations,'' in \emph{2005 18th International Conference on Applied
  Electromagnetics and Communications}, 2005, pp. 1--8.

\bibitem{10005237}
Y.~Huang, P.~Karadimas, and A.~Pour~Sohrab, ``Spatial channel degrees of
  freedom for optimum antenna arrays,'' \emph{IEEE Trans. Wireless Commun.},
  vol.~22, no.~8, pp. 5129--5144, Aug. 2023.

\bibitem{9373344}
W.~Ryoo and W.~Sung, ``Beamforming using uniform spherical arrays: Array
  construction, beam characteristics, and multi-rank transmission,'' \emph{IEEE
  Access}, vol.~9, pp. 38\,731--38\,741, Mar. 2021.

\bibitem{6905568}
O.~Franek and G.~F. Pedersen, ``Spherical arrays for wireless channel
  characterization and emulation,'' in \emph{2014 IEEE-APS Topical Conference
  on Antennas and Propagation in Wireless Communications (APWC)}, 2014, pp.
  480--483.

\bibitem{10817230}
Y.~Guo, Y.~Zhang, L.~Pang, Y.~Liu, and Z.~Ding, ``Non-uniform {3D} massive mimo
  arrays topology optimization for near-field communications,'' in \emph{2024
  IEEE 35th International Symposium on Personal, Indoor and Mobile Radio
  Communications (PIMRC)}, 2024, pp. 1--5.

\bibitem{10964237}
H.~Jing, W.~Cheng, and W.~Zhang, ``Spherical {RIS}-assisted mmwave {MIMO}
  wireless communications with concentric {UCAs},'' \emph{IEEE Trans. Commun.},
  vol.~73, no.~10, pp. 9676--9688, Oct. 2025.

\bibitem{famoriji2021source}
O.~J. Famoriji and T.~Shongwe, ``Source localization of {EM} waves in the
  near-field of spherical antenna array in the presence of unknown mutual
  coupling,'' \emph{Wirel. Commun. Mob. Comput.}, vol. 2021, no.~1, p. 3237219,
  2021.

\end{thebibliography}

\end{document}